\documentclass[twocolumn,showpacs,preprintnumbers,amsmath,amssymb]{revtex4}
\usepackage{graphicx}% Include figure files
\usepackage{dcolumn}% Align table columns on decimal point
\usepackage{bm}% bold math

\begin{document}

%\preprint{APS/123-QED}

\title{Secure Key Distribution by Swapping Quantum Entanglement}% Force line breaks with \\

\author{Daegene Song}%
\affiliation{%
National Institute of Standards and Technology\\ 100 Bureau Drive, MS 8910, Gaithersburg, MD 20899}%

\begin{abstract}
We report two key distribution schemes achieved by swapping quantum entanglement.
Using two Bell states, two bits of secret key can be shared between two distant parties
that play symmetric and equal roles.
We also address eavesdropping attacks against the schemes.

\end{abstract}

\maketitle
Cryptography has been one of the most fruitful applications coming out of quantum information theory and
it appears to be practically implementible in the nearest future among quantum
technology \cite{review}.
Since the first key distribution protocol using four quantum states was proposed in 1984 (called BB84) \cite{BB84},
a number of cryptographic methods based on quantum mechanics have been
proposed \cite{ekert,bennett2,goldenberg,bruss,bech,cabello1,cabello2,long,guo,beige,zhao,kim1,kim2}. At the heart of
quantum technology, including that of cryptography, lies entanglement.
Quantum entanglement is a subtle nonlocal correlation between the parts of a
quantum system and has no classical analog.
In 1991, Ekert showed \cite{ekert} that quantum
entanglement can be useful in sharing private keys between the two
parties. Suppose Alice and Bob share many maximally entangled
pairs of qubits. They then make measurements in jointly determined
random bases. After the measurements, Alice and Bob publicly
announce which basis they have used.
If they had measured in the same basis, the keys would
be perfectly correlated. Instead of discarding the keys resulting
from measuring in different bases, Alice and Bob use them to check
whether or not Bell's inequality is satisfied. If it is, then
Eve's presence is detected.  If not, Eve is absent and they keep
the perfectly correlated keys.
Entanglement swapping \cite{zukowski} (also see \cite{bose,hardy}) is
a method that enables one to entangle two quantum systems that do not have direct interaction with one another.
Based on entanglement swapping, quantum key distribution (QKD) protocols have been introduced \cite{cabello1,zhao,kim2} as well as
the ones without alternative measurements \cite{cabello1,cabello2,goldenberg,beige,zhao}
as seen in BB84 and Ekert's protocols.

In this paper, we report two QKD schemes using entanglement swapping which also
do not require alternative measurements, thereby improving the rate of generated key bits
per transmitted qubit, i.e. two bits per two Bell states.  This rate of generated key bits is an improvement from
the protocols introduced in \cite{cabello1,zhao}.
In order to illustrate entanglement swapping, we first define four Bell
states as $\Phi^{\pm}\equiv  (|00\rangle \pm |11\rangle )/\sqrt{2}$ and
$\Psi^{\pm} \equiv (|01\rangle \pm |10\rangle )/\sqrt{2}$.
Suppose two distant parties, Alice and Bob, share $\Phi_{12}^+$ and
$\Phi_{34}^+$ where Alice has qubits 1 and 4, and Bob possesses 2 and 3.
A measurement is performed on qubits 2 and 3 with the Bell basis, $\Phi^{\pm}$ and
$\Psi^{\pm}$,   then the total state is projected onto $|\eta_1\rangle = \Phi_{23}^+ \otimes \Phi_{14}^+$,
$|\eta_2\rangle = \Phi_{23}^- \otimes \Phi_{14}^-$, $|\eta_3\rangle = \Psi_{23}^+ \otimes \Psi_{14}^+$,
and $|\eta_4\rangle = \Psi_{23}^- \otimes \Psi_{14}^-$ with equal probability of 1/4 for each.
Previous entanglement between qubits 1 and 2, and 3 and 4 are now swapped into
entanglement between qubits 2 and 3, and 1 and 4.  Although we considered entanglement
swapping with the initial state $\Phi_{12}^+ \otimes \Phi_{34}^+$, similar results
can be achieved with other Bell states.  For example, when Alice and Bob initially
share $\Phi_{12}^- $ and $\Psi_{34}^+$, there are four possible measurement
outcomes with equal probability. If Bob gets $\Phi^+$ when qubits 2 and 3 are
measured, then Alice will obtain $\Psi^-$ for qubits 1 and 4.  We denote this
possibility as $\{ \Psi_{14}^-,\Phi_{23}^+\}$.  There are three other possibilities,
$\{ \Phi^{+}_{14}, \Psi^{-}_{23}\}$,  $\{ \Phi^{-}_{14}, \Psi^{+}_{23}\}$ and
$\{ \Psi^{+}_{14}, \Phi^{-}_{23}\}$.   Table \ref{cry1} shows Bell measurement
outcomes for initial states of a different combination of four Bell states.

\begin{figure}
\begin{center}
\includegraphics[scale=.5]{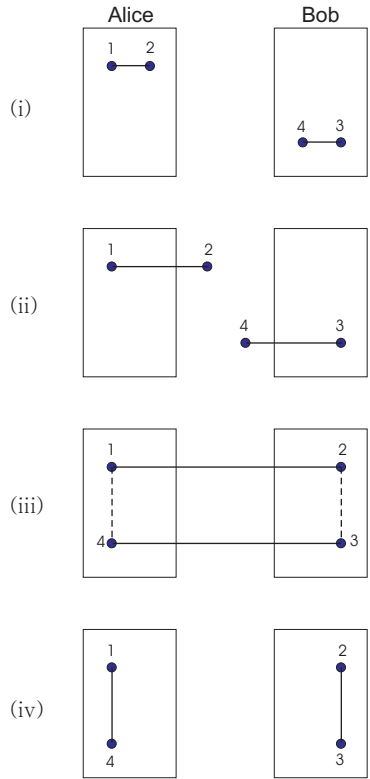}% Here is how to import EPS art
\caption{QKD using entanglement swapping.
The bold line means qubits are entangled and the dotted line
implies a Bell measurement is performed on the qubits.
(i) Alice and Bob each prepare arbitrary states, either in Bell or on a rotated bases, known only to themselves.
(ii) Each sends one qubit of the states to the other party.
(iii) After they confirm publicly the other person received the qubit and announce the choice of basis for initial preparation,
Alice and Bob perform Bell measurements on 1 and 4, and 2 and 3, respectively.
(iv) Alice and Bob now announce which Bell states had been prepared
(or their measurement result for $Scheme \, II$),
and they are able to find out the correlated measurement results
(or initial state preparation for $Scheme \, II$). }
\label{bbasic}
\end{center}
\end{figure}

In order to illustrate QKD based on entanglement
swapping,  we assign two bits to the measurement results as shown in
table \ref{bits}. This assignment of two bits has a stipulation that
one has to know both Alice and Bob's results in order to know the key
bits.  For instance, even if Eve knows $\Phi_{14}^+$ without knowing Bob's
result, there are still four different possible keys.
We also introduce another basis, i.e. a rotated basis,
$\omega^{\pm} \equiv  (|0+\rangle \pm |1-\rangle )/\sqrt{2} $ and
$\chi^{\pm} \equiv (|0-\rangle \pm |1+\rangle )/\sqrt{2} $
where $|+\rangle \equiv \sqrt{1/4}|0\rangle + \sqrt{3/4}|1\rangle $ and
$|-\rangle \equiv \sqrt{3/4}|0\rangle - \sqrt{1/4}|1\rangle $.
The first proposed QKD protocol, which we will call $Scheme \, I$,
goes as follows (see Fig. \ref{bbasic}):
\begin{enumerate}
\item[(S1)]
Alice prepares qubits 1 and 2 either in Bell basis, i.e.
chosen from $ \Psi^{\pm}_{12}$, $\Phi^{\pm}_{12}$, or on a rotated basis, $\omega_{12}^{\pm}, \chi_{12}^{\pm}$,
known only to herself.
\item[(S2)]
Bob also prepares qubits 3 and 4 either in Bell basis chosen from $ \Psi^{\pm}_{34}, \Phi^{\pm}_{34}$,
or on a rotated basis of $\omega_{34}^{\pm}, \chi_{34}^{\pm}$ , known only to
himself.
\item[(S3)]
 Alice sends qubit 2 to Bob and Bob transmits qubit 4 to Alice through public channels.
\item[(S4)]
Alice and Bob each publicly confirm that the other received the qubits.
\item[(S5)]
Alice and Bob also announce which basis have been used (i.e. either Bell basis or a rotated basis).
\item[(S6)]
If the received qubit had been prepared in the rotated basis,
Alice (or Bob) rotates back the received qubit into Bell basis by applying
$U_R=\sqrt{1/4}|0\rangle\langle 0| + \sqrt{3/4}|0\rangle\langle 1| +
\sqrt{3/4}|1\rangle\langle 0| -  \sqrt{1/4}|1\rangle\langle 1|$.
\item[(S7)]
Alice and Bob perform Bell measurements on 1 and 4, and 2 and 3, respectively.
\item[(S8)]
Alice and Bob announce which initial Bell states (If the initial
state were prepared on a rotated basis,
then it would correspond to one of four Bell states
after rotating it back, using the same
local unitary operation $U_R$ in (S6)) each had prepared.
\item[(S9)]
Now knowing the initial Bell states prepared by each other and their own measurement result, they could determine
which Bell measurement result the other had obtained.
\item[(S10)] Finally, with the given Bell measurement
result, Alice and Bob share two key bits according to table \ref{bits}.
\end{enumerate}
For example, suppose Alice initially prepared $\Psi_{12}^+$ (or $\chi^+_{12}$) while Bob
prepared $\Phi^-_{34}$ (or $\omega^-_{34}$).  After they publicly announce their prepared
states as in (S8), from table \ref{cry1}, they could find out that
there are four possible Bell measurement outcomes: $\{ \Phi^+_{14},\Psi^-_{23}\}$, $\{ \Phi^-_{14},\Psi^+_{23}\}$, $\{ \Psi^+_{14},\Phi^-_{23}\}$, and $\{ \Psi^-_{14},\Phi^+_{23}\}$.
Therefore, Alice and Bob, knowing their own measurement results,
could determine which correlated measurement result
both are sharing.  If Alice got $\Phi_{14}^-$ and Bob obtained $\Psi_{23}^+$,
then they will share the key bits \lq\lq 01\rq\rq   according to table \ref{bits}.

\begin{table}

\begin{center}

\begin{tabular}{|l||l|l|l|l|} \hline
  & $\Phi^+_{34}$ & $\Phi^-_{34}$ & $\Psi^+_{34}$ & $\Psi^-_{34}$ \\ \hline\hline
$\Phi^+_{12}$ & ${\bf {Id}}$ & $+-$ & $\Phi\Psi$ & both \\
$\Phi^-_{12}$ & $+-$ & ${\bf {Id}}$ & both & $\Phi\Psi$  \\
$\Psi^+_{12}$ &$ \Phi\Psi$ &  both & ${\bf {Id}}$ & $+-$ \\
$\Psi^-_{12}$ & both & $\Phi\Psi$ & $+-$ & ${\bf {Id}}$ \\ \hline
\end{tabular}

\end{center}
\caption{Alice (on qubits 1 and 4) and Bob's (on qubits 2 and 3) Bell measurement results for
 Bell states 1 and 2, and 3 and 4.  The symbol,
\lq\lq ${\bf {Id}}$\rq\rq means there are four possible outcomes for Alice and Bob's Bell measurement,
$\{ \Phi^{+}_{14}, \Phi^{+}_{23}\}$, $\{ \Phi^{-}_{14}, \Phi^{-}_{23}\}$,  $\{ \Psi^{+}_{14}, \Psi^{+}_{23}\}$ and
$\{ \Psi^{-}_{14}, \Psi^{-}_{23}\}$ with equal probability of 1/4.
Similarly, \lq\lq $+-$\rq\rq $\Rightarrow \{\Phi^{+}_{14}, \Phi^{-}_{23}\}$, $\{\Phi^{-}_{14}, \Phi^{+}_{23}\}$,
$\{\Psi^{+}_{14}, \Psi^{-}_{23}\}$, $\{\Psi^{-}_{14}, \Psi^{+}_{23}\}$,
and \lq\lq$\Phi\Psi$\rq\rq  $\Rightarrow$ $\{\Phi^{+}_{14}, \Psi^{+}_{23}\}$, $\{ \Phi^{-}_{14}, \Psi^{-}_{23}\}$,
$\{\Psi^{+}_{14}, \Phi^{+}_{23}\}$, $\{\Psi^{-}_{14}, \Phi^{-}_{23}\}$,
and \lq\lq both\rq\rq $ \Rightarrow$ $\{\Phi^{+}_{14}, \Psi^{-}_{23}\}$,
$\{\Phi^{-}_{14}, \Psi^{+}_{23}\}$, $\{ \Psi^{+}_{14}, \Phi^{-}_{23}\}$, $\{ \Psi^{-}_{14}, \Phi^{+}_{23}\}$.   }
\label{cry1}\end{table}

\begin{table}
\begin{center}

\begin{tabular}{|l||l|l|l|l|} \hline
  & $ {\bf {Id}}$ & $+-$ & $\Phi\Psi$ & both \\ \hline\hline
\lq\lq 00\rq\rq & $\{ \Phi_{14}^+,\Phi_{23}^+\}$ & $\{ \Psi_{14}^-,\Psi_{23}^+\}$ & $\{\Phi_{14}^-,\Psi_{23}^-\}$ & $\{ \Psi_{14}^+,\Phi_{23}^-\}$ \\
\lq\lq 01\rq\rq  & $\{ \Psi_{14}^-,\Psi_{23}^-\}$ & $\{ \Phi_{14}^+,\Phi_{23}^-\}$ & $\{\Psi_{14}^+,\Phi_{23}^+\}$ & $\{ \Phi_{14}^-,\Psi_{23}^+\}$  \\
\lq\lq 10\rq\rq  & $\{\Psi_{14}^+,\Psi_{23}^+\}$ & $\{\Phi_{14}^-,\Phi_{23}^+\}$ & $\{\Psi_{14}^-,\Phi_{23}^-\}$ & $\{\Phi_{14}^+,\Psi_{23}^-\}$ \\
\lq\lq 11\rq\rq  & $\{\Phi_{14}^-,\Phi_{23}^-\}$ & $\{\Psi_{14}^+,\Psi_{23}^-\}$ & $\{\Phi_{14}^+,\Psi_{23}^+\}$ & $\{\Psi_{14}^-,\Phi_{23}^+\}$ \\ \hline
\end{tabular}

\end{center}
\caption{Assignment of key bits for Alice and Bob's  measurement results for $Scheme \, I$.
In order to discover key bits, one has to know both Alice and Bob's measurement results.
For $Scheme \, II$, we assign two bits to the initial Bell state preparation,
thereby the subscripts 14 and 23 are replaced by 12 and 34, respectively, in the table.   }
\label{bits}\end{table}

In order to describe the second QKD scheme, we will assign two bits to the
initial state preparations.  Just as we assigned two bits to Alice and Bob's
measurement results, we will assign the same two bits where initial state for
qubits 1 and 2 replacing Alice's measurement result and initial Bell state of
qubits 3 and 4 replacing Bob's measurement result in table \ref{bits}.  As in $Scheme \, I$,
the key bits for $\omega^{\pm}$ ($\chi^{\pm}$)
will be same as $\Phi^{\pm}$ ($\Psi^{\pm}$).
The second protocol is very similar to the first one, and it proceeds the same as the first one
until (S7) of $Scheme \, I$. Then it goes as follows:
\begin{enumerate}
\item[(S$8^{\prime}$)] Alice
and Bob announce the result of their measurements.
\item[(S$9^{\prime}$)] Now each
knowing the measurement results for both and their own prepared state,
Alice and Bob could determine the initial state preparation by each other from table \ref{cry2}.
\item[(S$10^{\prime}$)] This enables Alice and Bob to share two key bits according to table \ref{bits}.
\end{enumerate}
For example, if Alice's measurement result is $\Phi_{14}^-$ and Bob obtains $\Psi_{23}^-$, then from table \ref{cry2},
there are four possible state preparation: $\{ \Phi^+_{12}, \Psi^+_{34} \}$,$\{ \Phi^-_{12}, \Psi^-_{34} \}$,
$\{ \Psi^+_{12}, \Phi^+_{34} \}$, and $\{ \Psi^-_{12}, \Phi^-_{34} \}$.
If Alice initially prepared $\Psi_{12}^+$ and Bob prepared $\Phi_{34}^+$
then they will share the key bits \lq\lq 01\rq\rq according to table \ref{bits}.

Although in the QKD schemes described above assert that Alice and Bob
publicly confirm whether the other received the qubits and announce
initial state preparation or measurement result, in practice,
this communication will be through a private channel.  In order to prevent
Eve from listening and altering these classical messages, encrypted messages
can be used.  Note that two proposed QKD schemes assume equal roles played by
Alice and Bob.

\begin{figure}
\begin{center}
\includegraphics[scale=.5]{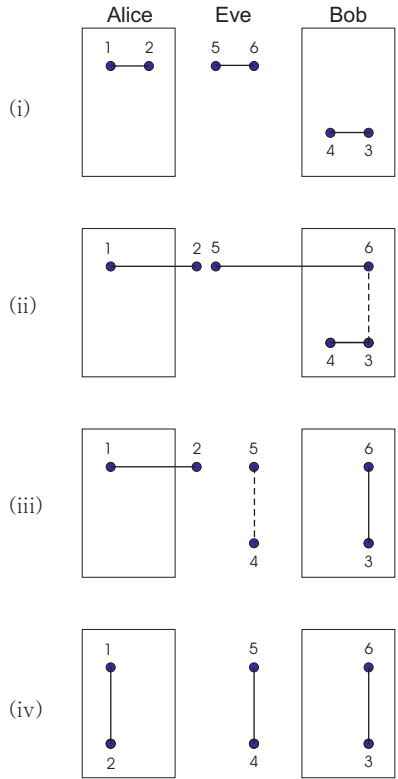}% Here is how to import EPS art
\caption{Eavesdropping scheme on the proposed entanglement swapping QKD.
As in Fig. \ref{bbasic}, the bold line means the qubits are entangled and the dotted line
implies a Bell measurement is performed on the qubits.
(i) While Alice and Bob prepare arbitrary states, Eve also prepares $\Phi_{56}^+$.
(ii) Eve intercepts qubit 2 sent by Alice and sends qubit 6 to Bob instead.
(iii) Eve performs a Bell measurement on qubits 5 and 4.
(iv) Eve then performs a local operation on qubit 2 and sends it back to Alice. }
\label{eve}
\end{center}
\end{figure}

As in BB84, Alice and Bob can detect eavesdropping by comparing the shared information publicly.
They will take out a sample and compare by publicly announcing
both the correlated measurement results and the initial states.
Comparing measurement result and initial states rather than key bits gives extra
safety since there are four different possible measurement results
(or initial preparations for $Scheme \, II$) for each key.
Let us consider an eavesdropping scenario for $Scheme\, I$ as shown in Fig. \ref{eve}.
Eve prepares $\Phi^+_{56}$ while Alice and Bob, as before, prepare initial states.
Alice sends her qubit 2, and Eve intercepts it and sends qubit 6 to Bob instead.
Suppose Alice and Bob use only the Bell basis rather than using both Bell and the rotated basis.
Eve could perform Bell measurement on qubits 5 and 4 and perform local unitary
operation, ${\bf {1}}$ for $\Phi_{54}^+$,
$\sigma_z$ for $\Phi_{54}^-$, $\sigma_x$ for $\Psi_{54}^+$, and $\sigma_x \sigma_z$ for $\Psi_{54}^-$,   on
qubit 2 and return it to Alice.  After Alice and Bob announce initial preparation,
Eve could find out the key bits Alice and
Bob are sharing.  However, this would not be possible since Alice and Bob prepared their initial states
in both Bell and the rotated bases and announced the choice of basis after they confirmed the other received the qubit.
Therefore, it is important for Alice and Bob not to reveal which state had been prepared initially and
perform measurements after they confirm the other received the qubits.

\begin{table}

\begin{center}

\begin{tabular}{|l||l|} \hline
 Measurement results & States prepared \\ \hline\hline
${\bf {Id}}$ & $ \Phi^+_{12} \Phi^+_{34}, \Phi^-_{12} \Phi^-_{34} , \Psi^+_{12} \Psi^+_{34}, \Psi^-_{12} \Psi^-_{34} $ \\
$+-$ &  $\Phi^+_{12} \Phi^-_{34}, \Phi^-_{12} \Phi^+_{34} , \Psi^+_{12} \Psi^-_{34}, \Psi^-_{12} \Psi^+_{34} $  \\
$\Phi\Psi$ &   $\Phi^+_{12} \Psi^+_{34}, \Phi^-_{12} \Psi^-_{34} , \Psi^+_{12} \Phi^+_{34}, \Psi^-_{12} \Phi^-_{34} $ \\
both &   $\Phi^+_{12} \Psi^-_{34}, \Phi^-_{12} \Psi^+_{34} , \Psi^+_{12} \Phi^-_{34}, \Psi^-_{12} \Phi^+_{34} $ \\ \hline
\end{tabular}

\end{center}
\caption{Possible state preparations for given measurement results for {\it {Scheme II}}.
When Alice and Bob obtain the measurement result as in the first column,
there are four possible initial Bell state preparation
(If the initial
state were prepared on a rotated basis,
then it would correspond to one of four Bell states after rotating it back,
using the same local unitary operation $U_R$ in (S6))   as shown in the second column.
The symbols in the\lq\lq Measurement results\rq\rq column
are the same as in Table \ref{cry1}.} \label{cry2}\end{table}

Practical feasibility of the proposed schemes can be sought in experiments that
use Bell operator measurements, such as teleportation \cite{teleportation}
and entanglement swapping \cite{pan}.  Successful Bell type measurements have been
performed using two photons, which were both
path- and polarization entangled.  Although there is an improvement of the key bit generation rate compared to
the protocol introduced by Zhao {\it {et. al.}} in \cite{zhao}, practical implementation of the proposed scheme will
have difficulty with classical information needed to perform neccessary
local unitary operations before performing a Bell measurement for each photon received.

The author thanks E. Knill for pointing out an error in the previous version of the proposed protocol and making
many helpful comments.  The author is also grateful to R. Boisvert, G. Brennen, and R. Kuhn for
helpful discussions.

\end{document}